\documentstyle[preprint,tighten,eqsecnum,aps,floats,psfig,epsfig,prb]{revtex}

\def\spose#1{\hbox to 0pt{#1\hss}}
\def\ltapprox{\mathrel{\spose{\lower 3pt\hbox{$\mathchar"218$}}
 \raise 2.0pt\hbox{$\mathchar"13C$}}}
\def\gtapprox{\mathrel{\spose{\lower 3pt\hbox{$\mathchar"218$}}
 \raise 2.0pt\hbox{$\mathchar"13E$}}}

\begin{document}
\draft
\title{
Instability of the O(5) multicritical behavior in the SO(5) theory of high-$T_c$
superconductors.
}
\author{Martin Hasenbusch$\,^1$, Andrea Pelissetto$\,^2$, 
Ettore Vicari$\,^1$ }
\address{$^1$
Dipartimento di Fisica dell'Universit\`a di Pisa 
and INFN, Pisa, Italy.
}
\address{$^2$ Dipartimento di Fisica dell'Universit\`a di Roma 
``La Sapienza" and INFN, Roma, Italy.}
\address{
{\bf e-mail: \rm 
{\tt Martin.Hasenbusch@df.unipi.it},
{\tt Andrea.Pelissetto@roma1.infn.it},
{\tt Ettore.Vicari@df.unipi.it}
}}

\date{\today}

\maketitle

\begin{abstract}
We study the nature of the multicritical point in the
three-dimensional O(3)$\oplus$O(2) symmetric Landau-Ginzburg-Wilson
theory, which describes the competition of two order parameters that
are O(3) and O(2) symmetric, respectively.  This study is relevant for
the SO(5) theory of high-$T_c$ superconductors, which predicts the
existence of a multicritical point in the temperature-doping phase
diagram, where the antiferromagnetic and superconducting transition
lines meet. 

We investigate whether the O(3)$\oplus$O(2) symmetry gets effectively
enlarged to O(5) approaching the multicritical point. For this
purpose, we study the stability of the O(5) fixed point. By means of a
Monte Carlo simulation, we show that the O(5) fixed point is unstable
with respect to the spin-4 quartic perturbation with the crossover
exponent $\phi_{4,4}=0.180(15)$, in substantial agreement with recent
field-theoretical results.  This estimate is much larger than the
one-loop $\epsilon$-expansion estimate $\phi_{4,4}=1/26$, which has
often been used in the literature to discuss the multicritical
behavior within the SO(5) theory. Therefore, no symmetry enlargement
is generically expected at the multicritical transition.

We also perform a five-loop field-theoretical analysis of the
renormalization-group flow. It shows that bicritical systems are not
in the attraction domain of the stable decoupled fixed point. Thus, in
these systems---high-$T_c$ cuprates should belong to this class---the
multicritical point corresponds to a first-order transition.
\end{abstract}



\section{Introduction}\label{sec1}

Multicritical phenomena arise from the competition of distinct types
of order.\cite{LF-72,FN-74,KNF-76,CPV-03} More specifically, a
multicritical point (MCP) is observed at the intersection of two
critical lines characterized by different order parameters.  MCPs are
expected in the temperature-doping phase diagram of high-$T_c$
superconductors, since, at low temperatures, these materials exhibit
both superconductivity and antiferromagnetism depending on doping.
The SO(5) theory \cite{Zhang-97,ZHAHA-99} attempts to provide a
unified description of these two phenomena by introducing a
three-component antiferromagnetic order parameter and a $d$-wave
superconducting complex order parameter.  This theory predicts a MCP
in the temperature-doping phase diagram where the two order parameters
become both critical.

Neglecting the fluctuations of the magnetic field and the quenched
randomness introduced by doping---see, e.g., Ref.~\CITE{Aharony-02-2}
for a critical discussion of this point---the critical behavior at the
MCP is determined by the O(3)$\oplus$O(2)-symmetric
Landau-Ginzburg-Wilson (LGW)
Hamiltonian,\cite{HZ-00,Hu-01,BL-98,AH-00,MN-00,KAE-01,DHZ-04} given
by
\begin{equation}
{\cal H} = \int d^d x \Bigl[ 
\case{1}{2} ( \partial_\mu \phi)^2  + \case{1}{2} (
\partial_\mu \psi)^2 + \case{1}{2} r_\phi \phi^2  
 + \case{1}{2} r_\psi \psi^2 
+ \case{1}{24}
u_0 (\phi^2)^2 + \case{1}{24}
v_0 (\psi^2)^2 + \case{1}{4} w_0 \,\phi^2\psi^2 \Bigr],
\label{bicrHH} 
\end{equation}
where $\phi$ and $\psi$ are respectively three- and two-component real
fields, associated with the antiferromagnetic and superconducting
order parameters, respectively.  The critical behavior at the MCP is
determined by the stable fixed point (FP) of the renormalization-group
(RG) flow when both $r_\phi$ and $r_\psi$ are tuned to their critical
value.  An interesting possibility, put forward in
Refs.~\CITE{Zhang-97,ZHAHA-99,HZ-00,Hu-01}, is that the
O(3)$\oplus$O(2) symmetry gets effectively enlarged to O(5) when
approaching the MCP. This requires the stability of the O(5) FP in the
theory (\ref{bicrHH}).

Cuprates have a pronounced two-dimensional layer structure with
relatively weak couplings between adjacent CuO$_2$ planes, so that
two- and three-dimensional models should represent two extreme
conditions for the possible range of properties of real high-$T_c$
superconductors, see, e.g., Ref.~\CITE{DHZ-04}.  In this paper we
investigate the stability properties of the O(5) FP in the
three-dimensional (3-$d$) multicritical theory (\ref{bicrHH}).

\begin{figure*}[tb]
\centerline{\psfig{width=8truecm,angle=0,file=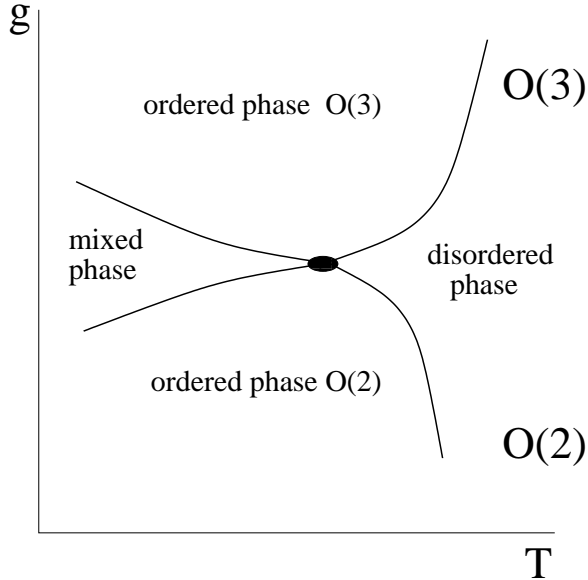}}
\vspace{2mm}
\caption{Phase diagram with a tetracritical point.
Here, $T$ is the temperature and $g$  a second relevant parameter.
}
\label{tetra}
\end{figure*}

\begin{figure}[tb]
\centerline{\psfig{width=8truecm,angle=0,file=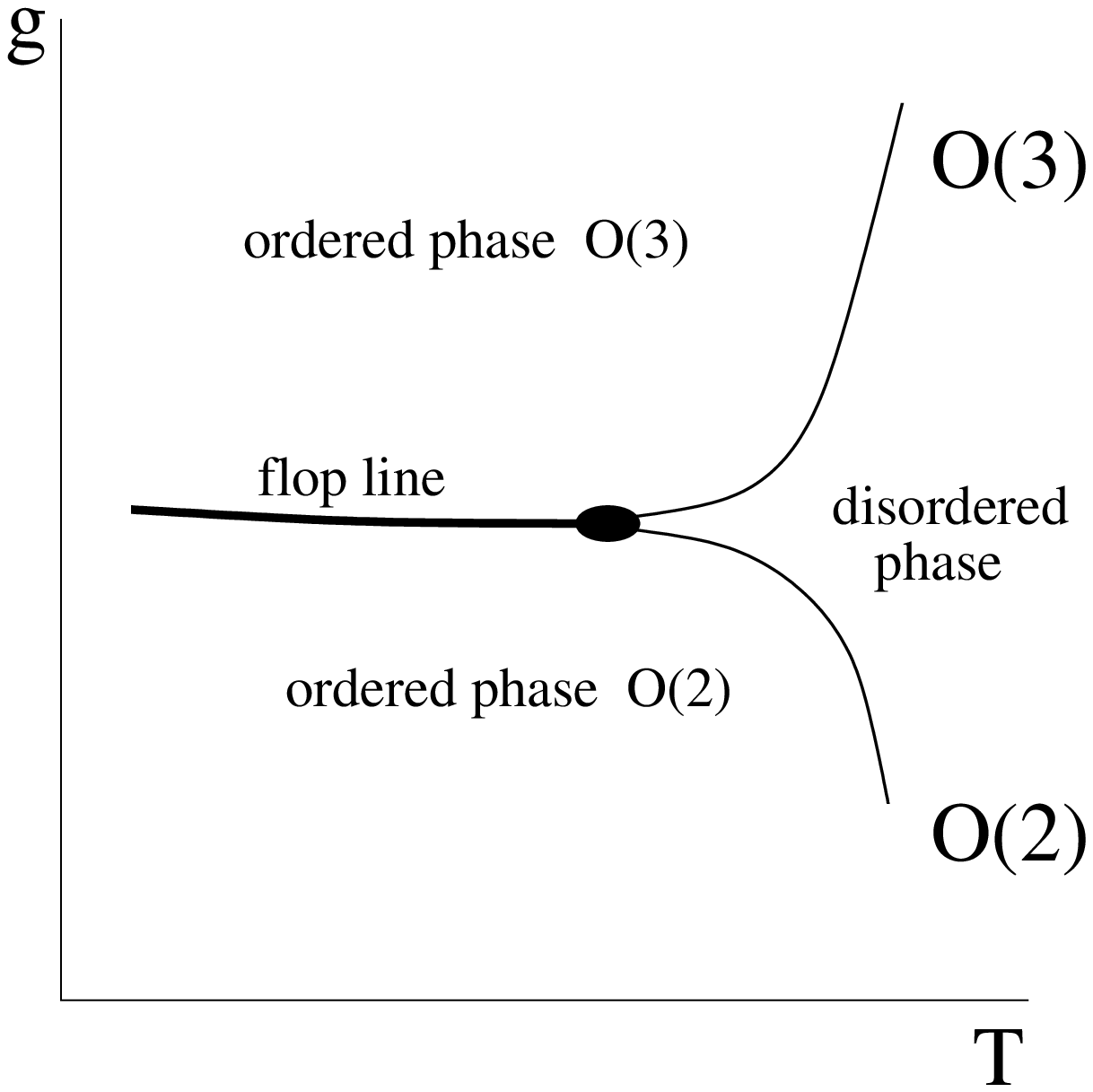}}
\vspace{2mm}
\caption{Phase diagram with a bicritical point.
Here, $T$ is the temperature and $g$  a second relevant parameter.
The thick line (``flop line") corresponds to a first-order transition.
}
\label{bicr}
\end{figure}

The phase diagram of the model with Hamiltonian (\ref{bicrHH}) has
been investigated within the mean-field approximation in
Ref.~\CITE{LF-72} (see also Ref.~\CITE{KAE-01}).  This analysis
predicts the existence of a bicritical or tetracritical point.  The
nature of the MCP depends on the sign of the quantity $\Delta_0\equiv
u_0 v_0 - 9 w_0^2$: If $\Delta_0 > 0$ the MCP is tetracritical as in
Fig.~\ref{tetra}, while for $\Delta_0<0$ it is bicritical, as in
Fig.~\ref{bicr}.

The nature of the MCP in the O(3)$\oplus$O(2) LGW theory has been
extensively studied.  In Refs.~\CITE{Zhang-97,HZ-00,Hu-01}, it was
speculated that the MCP is O(5)-symmetric.  In three dimensions, this
picture is supported by Monte Carlo (MC) simulations of a
five-component O(3)$\oplus$O(2)-symmetric spin model \cite{Hu-01} and
by a MC study of the quantum projected SO(5) model.\cite{JDAHZ-03}
These numerical studies showed that, within the parameter ranges
considered, the scaling behavior at the MCP is consistent with an
O(5)-symmetric critical behavior.  On the other hand, the one-loop
$\epsilon$-expansion results of Ref.~\CITE{KNF-76} show that the O(5)
FP is unstable.  Within this approximation, the stable FP is the
biconal FP, which has only O(3)$\oplus$O(2) symmetry.

The 3-$d$ RG flow of O($n_1$)$\oplus$O($n_2$) symmetric LGW theories
for generic $n_{1}$ and $n_2$ was investigated within the $\epsilon$
expansion to five loops.\cite{CPV-03} This analysis showed that the
O(5) FP and the biconal FP are both unstable. The stable FP is the
decoupled FP (DFP) corresponding to a multicritical behavior in which
the two order parameters $\phi$ and $\psi$ are effectively uncoupled.
The stability of the DFP can be proved by nonperturbative
arguments.\cite{Aharony-02,Aharony-02-2} Indeed, the RG dimension
$y_w$ of the operator $\phi^2 \psi^2$ that couples the two order
parameters is given by
\begin{equation}
y_w =  {1\over \nu_{\rm O(3)}} + {1 \over \nu_{\rm O(2)}} - 3,
\label{yw}
\end{equation}
where $\nu_{\rm O(3)}$ and $\nu_{\rm O(2)}$ are the critical exponents
of the O(3) and O(2) universality classes.  Using
\cite{CHPRV-02,CHPRV-01} $\nu_{\rm O(3)}=0.7112(5)$ and
$\nu_{\rm O(2)}=0.67155(27)$, one finds $y_w=-0.1048(12)$.  Note also
that the MCP described by the DFP is always tetracritical, as
in Fig.~\ref{tetra}.  No other stable FP is found within the 
$\epsilon$ expansion for the O(3)$\oplus$O(2) 
theory (\ref{bicrHH}).\cite{CPV-03,PV-04}  According to these RG results, the asymptotic
approach to the MCP within the SO(5) theory must be characterized by a
decoupled tetracritical behavior or by a first-order transition if
the system is outside the attraction domain of the DFP.

The analysis of Ref.~\CITE{CPV-03} indicates that the O(5) symmetry
can be asymptotically realized only by tuning an additional relevant
parameter, beside the double tuning required to approach the MCP. Of
course, if the system parameters are close to those needed to obtain
an O(5) MCP, one observes a crossover from the O(5) behavior to the
asymptotic one, which could be a decoupled critical behavior or a
first-order transition. This crossover can be described in terms of
universal scaling functions determined by the O(5) FP. For instance,
if $T_{O(5)}$ is the critical temperature of the O(5) MCP, $g_2$ and
$g_4$ are scaling fields (functions of the system parameters)
associated with the two relevant perturbations of the O(5) FP such
that $g_2 = g_4 = 0$ at the O(5) MCP, then the singular part of the
free energy can be written as
\begin{equation}
F_{\rm sing}= t^{d\nu} f(g_{2} t^{-\phi_{2,2}}, g_{4} t^{-\phi_{4,4}}) ,
\label{freen}
\end{equation}
where $t = (T - T_{O(5)})/T_{O(5)}$ is the reduced temperature.  Here
$\nu$ is the O(5) correlation-length exponent, $\phi_{2,2}>0$ and
$\phi_{4,4}>0$ are the crossover exponents associated with the two
relevant perturbations $g_{2}$ and $g_{4}$, which are determined by
the O(5) FP itself.

It has been noted \cite{Aharony-02,JDAHZ-03,DHZ-04} that the crossover
exponent $\phi_{4,4}$ associated with this additional instability of
the O(5) FP is rather small. This may partially explain the apparent
O(5) multicritical behavior observed in Refs.~\CITE{Hu-01,JDAHZ-03}:
The MC simulations are still probing a region in which the crossover
towards the asymptotic critical behavior is so slow to be
undetectable.  Within the $\epsilon$ expansion, one finds
$\phi_{4,4}=\frac{1}{26} \epsilon + O(\epsilon^2)$.  Using the value
suggested by setting $\epsilon=1$, i.e. $\phi_{4,4}\approx
1/26=0.038...$, the authors of Refs.~\CITE{JDAHZ-03,DHZ-04} argued
that any significant deviation away from the O(5)-symmetric FP can be
observed in experiments and realistic MC simulations only when the
reduced temperature is $t_{\rm cross}\approx 10^{-10}$, making the
departure away from the O(5) symmetric point practically unobservable.
On the other hand, the analysis of the perturbative expansions in two
different schemes, the 3-$d$ massive zero-momentum scheme (six loops)
and the $\epsilon$ expansion (five loops), gives a much larger
estimate of $\phi_{4,4}$,\cite{CPV-03} $\phi_{4,4}\approx 0.15$, which
makes the statement of Refs.~\CITE{JDAHZ-03,DHZ-04} on the effective
relevance of the O(5) FP very questionable.\cite{CPV-03}

In this paper we return to the issue of the stability of the O(5) FP.
We compute the crossover exponent $\phi_{4,4}$ using lattice
techniques based on MC simulations. We obtain $\phi_{4,4} =
0.180(15)$. This result is slightly larger than the field-theoretical
(FT) ones reported in Ref.~\CITE{CPV-03} and fully confirms the fact
that the naive application of the one-loop $\epsilon$-expansion result
gives an estimate that is unrealistically small.  Moreover, we perform
a detailed FT analysis of the RG flow in the 3-$d$ scheme known as
minimal-subtraction scheme without $\epsilon$ expansion.\cite{SD-89}
We find that bicritical systems that have $\Delta_0 < 0$ never flow
towards the DFP. Therefore, these systems, which are those of interest
for high-$T_c$ superconducting materials according to the analysis of
Ref.~\CITE{AH-00}, should have a first-order MCP.

The paper is organized as follows. In Sec.~\ref{sec2} we discuss the
stability of the O(5) FP and present a MC determination of the
crossover exponent $\phi_{4,4}$. In Sec.~\ref{sec3} we discuss the RG
flow and determine the attraction domain of the DFP.  In
Sec.~\ref{sec4} we report our conclusions. The five-loop perturbative
series used in Sec.~\ref{sec3} to determine the RG flow are reported
in the appendix.

\section{Stability of the O(5) fixed point} \label{sec2}

\subsection{General considerations} \label{sec2.1}

In order to determine the stability properties of the O(5) FP, we must
determine the RG dimensions of the perturbations present in the
Hamiltonian (\ref{bicrHH}) that break the O(5) symmetry down to ${\rm
O(3)}\oplus{\rm O(2)}$.  They can be expressed in terms of homogeneous 
polynomials of the fields $P_{m\l}$, where $m$ is the power of the fields and 
$\l$ the spin of the representation of the O(5) group.  The classification
in terms of spin values is particularly convenient, since polynomials
with different spin do not mix under RG transformations and the RG
dimension $y_{m\l}$ does not depend on the particular component of the
spin-$l$ representation.  We refer to Ref.~\CITE{CPV-03} for details.
For $m=2$ (resp. 4), the only possible values of $\l$ are $\l=0,2$
(resp. $\l=0,2,4$). Explicitly, we define
\begin{eqnarray}
&&P_{2,0}= \case{1}{2} ( \phi^2 + \psi^2), \nonumber \\
&&P_{2,2}= \case{2}{5} \phi^2-\case{3}{5} \psi^2 , \nonumber \\
&&P_{4,0}= P_{2,0}^{\,2},\nonumber \\
&&P_{4,2}= P_{2,0} P_{2,2}, \nonumber \\
&&P_{4,4}= \case{40}{63} \phi^2 \psi^2 - 
\case{5}{21} (\psi^2)^2 - \case{8}{63} (\phi^2)^2. 
\label{ppop} 
\end{eqnarray}
Note that in general all these operators renormalize multiplicatively
except $P_{4,2}$ that may mix with the lower-dimensional operator
$P_{2,2}$. 

Hamiltonian (\ref{bicrHH}) can be rewritten in the form
\begin{eqnarray}
{\cal H} &=& \int d^d x \left[
\case{1}{2} ( \partial_\mu \phi)^2  + \case{1}{2} (
\partial_\mu \psi)^2 + \case{1}{2} (r_0 P_{2,0} + r_2 P_{2,2}) + \right. 
\nonumber \\
&& \left. \qquad\qquad
+ \case{1}{24} (f_{0,0} P_{4,0} + f_{2,0} P_{4,2} + f_{4,0} P_{4,4}) \right],
\label{HO5} 
\end{eqnarray}
where
\begin{eqnarray}
&&f_{0,0} = \frac{4}{35} (15 u_0 + 8 v_0 + 36 w_0), \nonumber \\
&&f_{2,0} = \frac{4}{9} (5 u_0 - 4 v_0 - 3 w_0), \nonumber \\ 
&&f_{4,0} = 6w_0 - u_0 - v_0. 
\label{deffff}
\end{eqnarray}
If $r_2 = f_{2,0} = f_{4,0} = 0$ we obtain the O(5)-invariant
Hamiltonian.

The RG dimensions of the relevant quadratic perturbations $P_{2,0}$
and $P_{2,2}$ are respectively \cite{AS-95,CPV-03,exponents,CP-04}
$y_{2,0}=1/\nu = 1.31(1)$ ($\nu$ is the O(5) critical exponent
associated with the correlation length) and $y_{2,2}=1.83(1)$.  The
corresponding crossover exponents are $\phi_{2,0}=1$ and
$\phi_{2,2}=y_{2,2}\nu=1.40(2)$.  The corresponding Hamiltonian
parameters $r_\phi$ and $r_\psi$ must be tuned to approach the MCP.
The spin-0 quartic perturbation $P_{4,0}$ determines the leading
scaling corrections in the O(5) vector model; its RG dimension is
$y_{4,0}=-0.79(2)$.\cite{CPV-03} The spin-2 quartic perturbation
$P_{4,2}$ turns out to be irrelevant:
$y_{4,2}=-0.441(13)$.\cite{CPV-03} The O(5) FP is also unstable with
respect to the perturbation $P_{4,4}$ (more generally, the O($N$) FP
is unstable against the spin-4 quartic perturbation for $N>N_c$ with
$N_c\lesssim 3$, see Refs.~\CITE{review,KS-95,CH-98,CPV-00}).  The
corresponding RG dimension $y_{4,4}$ is given by
$y_{4,4}=\frac{1}{13}\epsilon+ O(\epsilon^2)$ close to four
dimensions.  A naive extrapolation of this result to three dimensions,
achieved by setting $\epsilon=1$, suggests a very small value.  On the
other hand, \cite{CPV-00,CPV-03,CP-04} a six-loop analysis of the
3-$d$ perturbative series in the massive zero-momentum scheme gives
$y_{4,4}=0.189(10)$, while the five-loop $\epsilon$ expansion gives
$y_{4,4}=0.198(11)$.  These estimates are still moderately small,
though much larger than the estimate $y_{4,4}=1/13$ obtained by
naively setting $\epsilon=1$ in the one-loop $\epsilon$-expansion result.

\subsection{Monte Carlo determination of the critical exponent $y_{4,4}$}

In the following we present a MC calculation of the RG dimension
$y_{4,4}$ and of the crossover exponent $\phi_{4,4} \equiv \nu
y_{4,4}$.  As we shall see, the result will provide conclusive
evidence for the fact that the 3-$d$ exponent $\phi_{4,4}$ is actually
much larger that the $O(\epsilon)$ result, in agreement with the
high-order FT analyses.

We consider the 5-vector model on a simple cubic lattice of size 
$L^3$ with Hamiltonian
\begin{equation}
{\cal H} = - \beta \sum_{\langle xy\rangle}   \vec{s}_{x}  \cdot \vec{s}_{y},
\label{lattO5}
\end{equation}
where $\vec{s}_x$ is a five-component unit vector, and 
the summation is extended over all nearest-neighbor pairs 
$\langle xy\rangle$.  The RG
dimension $y_{4,4}$ can be obtained by studying the scaling behavior
of the cubic-symmetric perturbation
\begin{equation}
P_c = \sum_x \sum_{i=1}^5  s_{x,i}^4
\label{cubicp}
\end{equation}
at the critical point.  Indeed, the operator $P_c$ 
is a particular combination of the spin-4 and spin-0
operators \cite{CPV-03} and thus its scaling behavior allows the determination
of $y_{4,4}$ (assuming of course that $y_{4,0} < y_{4,4}$).
We follow closely Ref.~\CITE{CH-98} where $y_{4,4}$ was computed
in the $N$-vector model for $N=2$, 3, and 4.  We define
\begin{equation}
\vec{M} \equiv \sum_x \vec{s}_x, 
\qquad R= \frac{ \sum_{i=1}^5 M_i^4 }{(M^2)^2},
\end{equation}
and compute the correlation
\begin{equation}
D_R \equiv \langle R P_c \rangle - \langle R \rangle \langle P_c \rangle\; .
\label{drdef}
\end{equation}
The RG dimension $y_{4,4}$ is obtained from the finite-size
scaling behavior of $D_R$ at the critical point, since
\begin{equation}
\label{powerlaw}
D_R(L,\beta=\beta_c) \sim L^{y_{4,4}}
\end{equation}
for $L\to \infty$.

In our simulations we used combinations of wall-cluster,\cite{HaPiVi99}
single-cluster,\cite{wolff} and overrelaxation
updates.  In order to determine the critical point and the
standard critical exponents, we considered lattices of size $L\le 128$ 
for several values of  $\beta$ close to 1.1813. 
We measured  the Binder cumulant
\begin{equation}
\label{Binderc}
 U_4 = \frac{\langle (\vec{M}^2)^2 \rangle}{ \langle \vec{M}^2 \rangle^2}  
\end{equation}
and the ratio $Z_a/Z_p$, where $Z_p$ is the partition function with
periodic boundary conditions and $Z_a$ is the partition function with
periodic boundary conditions in two directions and antiperiodic
boundary conditions in the third one (for a detailed discussion of
this quantity see Refs.~\CITE{Hasenbusch-01,CHPRV-01,CHPRV-02} and
references therein).  

The critical value 
$\beta_c$ is estimated by employing the standard crossing method for
the Binder cumulant and for $Z_a/Z_p$.  We obtain
\begin{equation}
\beta_c=1.18138(3),
\label{betac}
\end{equation}
where the quoted uncertainty includes both statistical and
systematical errors.  It is compatible with the estimate
$\beta_c=1.18127(10)$ obtained using only lattices with $L\le 32$,
indicating that scaling corrections are quite small.  The result
(\ref{betac}) is consistent with the estimate \cite{betac} obtained
from the analysis of the 21st-order high-temperature expansion of the
magnetic susceptibility computed in Ref.~\CITE{BC-97}, but much more
precise. The analysis also provides the critical values of $U_4$ and
of $Z_a/Z_p$: $U_4^* = 1.069(1)$ and $(Z_a/Z_p)^* = 0.071(1)$.

We estimate the exponent $\nu$
from the $L$-dependence of the slope of $Z_a/Z_p$ and of $U_4$ along
the line in the $(\beta, L)$ plane on which $Z_a/Z_p=0.071$ (which
means that, for each $L$, we take the two quantities at the value of
$\beta$ where $Z_a/Z_p$ takes the value $0.071$).
 Fitting the slope of $Z_a/Z_p$ for
data with $16\le L \le 128$ with a simple power-law Ansatz, we obtain
$\nu=0.7782(14)$ with $\chi^2/$d.o.f.$=0.49$, where d.o.f. indicates
the number of degrees of freedom of the fit.  If we use the data with
$8\le L \le 64$ we obtain instead $\nu=0.7766(5)$ with
$\chi^2/$d.o.f.$=0.95$.  The systematic error due to scaling
corrections should not be (much) larger than the difference between
these two results. Roughly, assuming that scaling corrections
are dominated by the leading one with exponent $\omega=-y_{4,0}=0.79(2)$,
the systematic error on $\nu=0.7782(15)$
should be given approximately by $(2^{\omega} - 1)^{-1}$ times the
difference, i.e. it should be approximately 0.002.  Fitting the slope
of the Binder cumulant, we obtain $\nu=0.7809(32)$
(resp. $\nu=0.7802(13)$) from data with $16\le L \le 128$ (resp. $8\le
L \le 64$).  These results are consistent with those from the slope of
$Z_a/Z_p$ but less precise.  As our final estimate we give
\begin{equation}
\nu=0.779(3), 
\label{nues}
\end{equation}
where the error should also include the systematic uncertainty due to scaling
corrections.  This estimate clearly rules out the MC result
$\nu=0.728(18)$ of Ref.~\CITE{Hu-01}.  It is also slightly larger than
the FT estimate \cite{exponents} $\nu=0.762(7)$.  Moreover, by fitting
the magnetic susceptibility $\chi$ at $Z_a/Z_p=0.071$ we find
\begin{equation}
\eta=0.034(1), 
\label{etaes}
\end{equation}
where again the quoted error includes possible systematic errors due
to scaling corrections.  Performing this analysis we followed closely
Refs.~\CITE{Hasenbusch-01,CHPRV-01,CHPRV-02}, where the models with 2,
3, and 4 components were studied.

The determination of $y_{4,4}$ through Eq.~(\ref{powerlaw}) requires a
very large statistics, which limits us to lattices with $L\le 24$.
Typically, approximately $10^9$ measurements were performed for each lattice
size.  Most of the simulations were performed at $\beta=1.18127$. In
order to extrapolate the data to other values of $\beta$ and, in
particular, to $\beta_c$, we also performed simulations at
$\beta=1.18$ and $\beta=1.183$, obtaining an estimate of the
derivative of $D_R$ with respect to $\beta$. In Table
\ref{rawdata} we summarize the numerical results for $D_R$ at
$\beta=1.18138\approx \beta_c$.

\begin{table}
\caption{\label{rawdata}
Estimates of $D_R$ at $\beta=1.18138$. The number reported 
in parentheses is the statistical error, while the number in brackets 
gives the error due to the uncertainty on the estimate of $\beta_c$. 
}
\begin{center}
\begin{tabular}{rl}
\multicolumn{1}{c}{$L$}&
\multicolumn{1}{c}{$D_R$}\\
\hline
        4 & 0.023979(6)[4] \\
        5 & 0.025668(8)[6] \\
        6 & 0.026950(10)[9]\\
        7 & 0.028019(13)[10] \\
        8 & 0.028942(16)[12] \\
        9 & 0.029733(18)[15] \\
       10 & 0.030501(22)[18] \\
       11 & 0.031100(25)[22] \\
       12 & 0.031752(28)[25] \\
       14 & 0.032900(36)[31] \\
       16 & 0.033825(44)[38] \\
       18 & 0.034769(53)[47] \\
       20 & 0.035638(81)[47] \\
       22 & 0.036461(72)[60] \\
       24 & 0.037134(88)[70] \\
\end{tabular}
\end{center}
\end{table}

First, we fit these data with the simple power-law
ansatz~(\ref{powerlaw}).  The results are summarized in Table
\ref{simplefit}.  The fits are stable starting from $L_{\rm
min}=8$. Also the $\chi^2/$d.o.f. for fits with $L_{\rm min}\ge 8$ is
smaller than one.
In order to check the dependence of the result for $y_{4,4}$ on the
numerical estimate of $\beta_c$, we repeat these fits using the
values of $D_R$ at $\beta=1.18135$ and $\beta=1.18141$, which
correspond to adding or subtracting one error bar to $\beta_c$.  For
example, for $L_{\rm min}=11$ we obtain the estimates
$y_{4,4}=0.2256(19)$ and $y_{4,4}=0.2283(19)$ for $\beta=1.18135$ and
$\beta=1.18141$, respectively. Optimistically, we could quote the
final result $y_{4,4}=0.227(3)$, where the error bar includes the
statistical uncertainty as well as the error due to the uncertainty in
the estimate of $\beta_c$. This estimate would be reliable if scaling
corrections, and in particular the leading ones associated with the
exponent $\omega=-y_{4,0}=0.79(2)$, are suppressed. This
seems to be supported by the
above-reported analysis to determine $\beta_c$ and the critical 
exponents, where no evidence for the leading scaling corrections 
was found within our statistical errors.

\begin{table}
\caption{\label{simplefit}
Fits of $D_R(\beta = 1.18138\approx \beta_c)$ on lattices 
of size $L\ge L_{\rm min}$
with the power-law Ansatz $D_R = c L^{y_{4,4}}$. 
In the final column
the $\chi^2$ per degree of freedom (d.o.f.) is reported.
}
\begin{center}
\begin{tabular}{rlll}
\multicolumn{1}{c}{$L_{\rm min}$}&
\multicolumn{1}{c}{$y_{4,4}$} &
\multicolumn{1}{c}{$c$} &
\multicolumn{1}{c}{$\chi^2/$d.o.f.}\\
\hline
    6   &  0.2333(3)  & 0.01778(2)  &     8.30 \\
    7   &  0.2290(8)  & 0.01796(3)  &     2.20 \\
    8   &  0.2268(10) & 0.01807(4)  &     0.93 \\
    9   &  0.2261(12) & 0.01810(6)  &     0.92 \\
   10   &  0.2252(15) & 0.01815(7)  &     0.90 \\
   11   &  0.2270(19) & 0.01806(9)  &     0.60 \\
\end{tabular}
\end{center}
\end{table}

In order to be conservative, we also analyze the data allowing for a
nonvanishing scaling correction, i.e. we fitted our data with the
ansatz
\begin{equation}
\label{corrections}
D_R(L,\beta=\beta_c) =c  L^{y_{4,4}} \; (1 + c' L^{-\omega})  \;\;,
\end{equation}
where we set $\omega=0.79$.\cite{CPV-03} Note that the
results for $y_{4,4}$ depend very little on the precise value 
for $\omega$.  The results of the fits are summarized in Table
\ref{correctionfit}.  Only for $L_{\rm min}<8$ do we get an amplitude
$c'$ that is different from zero within error bars. However, for these
fits also $\chi^2/$d.o.f. is large. This suggests that for the small
lattice sizes $L<8$ actually subleading scaling corrections dominate.
The results for the exponent $y_{4,4}$ obtained from $L_{\rm min}>7$
are completely consistent with the results reported above for the
power-law ansatz without corrections. However, here the error bars are
considerably larger.  Since we are not able to exclude leading
corrections to scaling on a firm basis, we take the error resulting
from the fits that include leading corrections to scaling as our final
estimate. Taking into account also the uncertainty in our estimate of
$\beta_c$ we arrive at the final estimate
\begin{equation}
y_{4,4}=0.23(2). 
\label{y44es}
\end{equation}
Note that the reported error estimate is probably quite conservative.
Using the estimate (\ref{nues}) of $\nu$, we obtain 
$\phi_{4,4}\equiv y_{4,4}\nu= 0.180(15)$ for the
corresponding crossover exponent.

\begin{table}
\caption{\label{correctionfit}
Fits of $D_R(\beta = 1.18138\approx \beta_c)$ on lattices 
of size $L\ge L_{\rm min}$
with the Ansatz (\ref{corrections}). The exponent $\omega$ was fixed: 
$\omega=0.79$.
In the final column
the $\chi^2$ per degree of freedom (d.o.f.) is reported.
}
\begin{center}
\begin{tabular}{rllll}
\multicolumn{1}{c}{$L_{\rm min}$}&
\multicolumn{1}{c}{$y_{4,4}$} &
\multicolumn{1}{c}{$c$} &
\multicolumn{1}{c}{$c'$} &
\multicolumn{1}{c}{$\chi^2/$d.o.f.}\\
\hline
 5  &  0.189(3)  & 0.0208(2) &          --0.321(16) & 3.02  \\
 6  &  0.199(4)  & 0.0201(3) &          --0.244(27) & 1.87  \\
 7  &  0.210(5)  & 0.0193(3) &          --0.152(37) & 1.24  \\
 8  &  0.221(8)  & 0.0185(6) &          --0.054(70) & 0.98  \\
 9  &  0.226(10) & 0.0181(7) &          --0.001(95) & 1.06  \\
10  &  0.245(15) & 0.0168(10)& \phantom{+}0.20(16)  & 0.83  \\
11  &  0.226(18) & 0.0182(13)&          --0.01(20)  & 0.71  \\
\end{tabular}
\end{center}
\end{table}

\section{Renormalization-group flow} \label{sec3}

The presence of a stable FP does not imply that the MCP should always
correspond to a second-order transition controlled by the stable
FP. Indeed, this happens only if the system is in the attraction
domain of the stable FP.  It is thus of interest to determine the
attraction domain of the DFP. For this purpose we study the RG flow of
the system, i.e., the RG trajectories along which the quartic
Hamiltonian parameters $u_0$, $v_0$, and $w_0$ are kept fixed.  

We consider the flow directly in three dimensions, using the
minimal-subtraction ($\overline{\rm MS}$) scheme without $\epsilon$
expansion.\cite{SD-89} In this scheme one considers the massless
(critical) theory in dimensional regularization within the
$\overline{\rm MS}$ scheme.~\cite{tHV-72} RG functions are obtained in
terms of the renormalized couplings $u,v,w$ and of $\epsilon \equiv 4
- d$. Subsequently $\epsilon$ is set to its physical value $\epsilon =
1$, providing a 3-$d$ scheme in which the 3-$d$ RG
functions are expanded in powers of the $\overline{\rm MS}$
renormalized quartic couplings $u,v,w$. This scheme differs from the
standard $\epsilon$ expansion~\cite{WF-72} in which one expands the RG
functions in powers of $\epsilon$.  The $\overline{\rm MS}$ $\beta$
functions have been computed to five loops for generic
O($n_1$)$\oplus$O($n_2$) LGW theories.~\cite{CPV-03} In the appendix
we report the five-loop series that are used in this section to
determine the RG flow of the O(3)$\oplus$O(2) case.

The RG trajectories are obtained by solving the differential equations
\begin{eqnarray}
&&- \lambda {du\over d\lambda} = \beta_u[u(\lambda),v(\lambda),w(\lambda)],
\nonumber \\
&&  - \lambda {dv\over d\lambda} = \beta_v[u(\lambda),v(\lambda),w(\lambda)],
\nonumber \\
&&  - \lambda {dw\over d\lambda} = \beta_w[u(\lambda),v(\lambda),w(\lambda)],
\end{eqnarray}
with $\lambda\in [0,\infty)$ and the initial conditions
\begin{eqnarray}
   &&u(0) = v(0) = w(0) = 0, \nonumber \\
   && \left. {du\over d\lambda}\right|_{\lambda=0} = u_0, \qquad
   \left. {dv\over d\lambda}\right|_{\lambda=0} = v_0, \qquad
   \left. {dw\over d\lambda}\right|_{\lambda=0} = w_0.
\end{eqnarray}
Note that the trajectories do not depend on the Hamiltonian parameters
individually, but only through their dimensionless ratios. For
instance, by rescaling $\lambda \to \lambda/w_0$, the initial
conditions depend only on $u_0/w_0$ and $v_0/w_0$.  The necessary
resummation of the series is perfomed by employing the Pad\'e-Borel
method; see, e.g., Refs.~\CITE{ZJbook,review}.

\begin{figure}[tb]
\centerline{\psfig{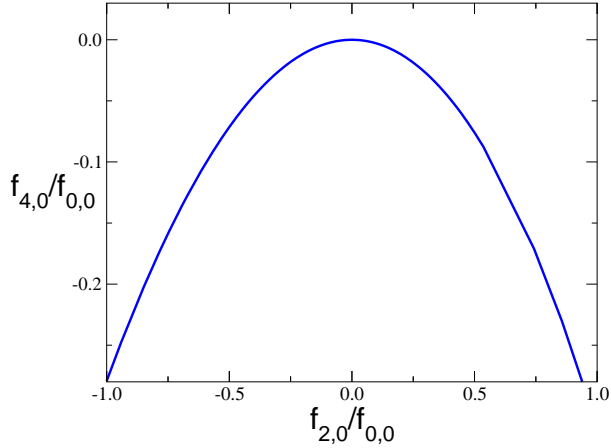}}
\vspace{2mm}
\caption{Initial conditions for trajectories flowing to the
O(5) FP in the $f_{2,0}/f_{0,0}$, $f_{4,0}/f_{0,0}$ plane.
}
\label{flowf2f4}
\end{figure}

\begin{figure}[tb]
\centerline{\psfig{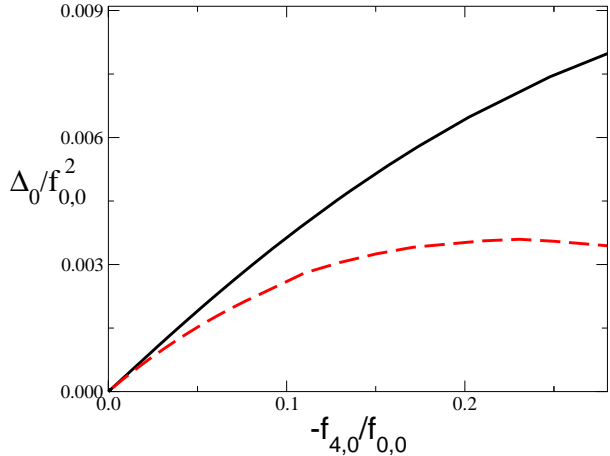}}
\vspace{2mm}
\caption{Initial conditions for trajectories flowing to the 
O(5) FP in the $f_{4,0}/f_{0,0}$, $\Delta_0/f_{0,0}^2$ plane.
Points belonging to the lower dashed curve have $f_{2,0}/f_{0,0} > 0$, 
those belonging to the upper continuous one have $f_{2,0}/f_{0,0} < 0$.
}
\label{flowf4Delta}
\end{figure}

In this RG scheme the coordinates of the O(5) FP are
$u^*=v^*=3w^*\approx 0.62$.  Since the O(5) FP has only one
instability direction in the $u$, $v$, $w$ space, RG trajectories flowing
to the O(5) FP separate the flow into two different domains.  
In order to characterize these domains, we consider the dimensionless
ratios $f_{2,0}/f_{0,0}$, $f_{4,0}/f_{0,0}$, cf. Eq.~(\ref{deffff}),
and $\Delta_0/f_{0,0}^2$ where $\Delta_0=u_0 v_0 - 9 w_0^2$.  
In Figs. \ref{flowf2f4} and \ref{flowf4Delta} we report the initial
conditions that correspond to trajectories flowing to the O(5) FP,
respectively in terms of $f_{2,0}/f_{0,0}$ and $f_{4,0}/f_{0,0}$
and in terms of $\Delta_0/f_{0,0}^2$ and $f_{4,0}/f_{0,0}$.  The
results reported in the figures were obtained by resumming the $\beta$
functions with a [5/1] Pad\'e-Borel approximant. We used several
different values of the Borel-Leroy parameter $b$: the results are
essentially independent of this parameter.

Since the spin-4 perturbation is relevant, asymptotically the
RG trajectories reaching the O(5) FP lie in the plane $f_{4} = 0$. One
may then naively expect that the initial conditions for trajectories
flowing to the O(5) FP have $f_{4,0}/f_{0,0}$ small. As it can be seen
from Fig.~\ref{flowf2f4} this is not really true, although
$|f_{4,0}/f_{0,0}| \ll |f_{2,0}/f_{0,0}|$. We also report the results
in terms of $\Delta_0/f_{0,0}^2$ for two reasons. First, $\Delta_0
\equiv u_0 v_0 - 9 w_0^2$ determines the nature (bicritical or
tetracritical) of the phase diagram. Second, in the mean-field
approximation $\Delta_0 = 0$ is required for obtaining an
O(5)-invariant transition.  Indeed, if one ignores fluctuations and
thus neglects the kinetic term and rescales the fields as $\phi \to
(\bar{u}/u_0)^{1/4} \phi$ and $\psi \to (\bar{u}/v_0)^{1/4} \psi$, one
obtains an O(5)-invariant quartic potential only for $\Delta_0 = 0$.
Note that, as show by Fig.~\ref{flowf4Delta}, this mean-field
condition is not actually satisfied when fluctuations are taking into
account, although $\Delta_0/f_{0,0}^2$ is small whenever
$f_{4,0}/f_{0,0}$ is also small.

The curve reported in Fig.~\ref{flowf2f4}, of equation
$f_{4,0}/f_{0,0} = g(f_{2,0}/f_{0,0})$, separates the space of initial
conditions into two different domains. We determine now which of them
contains the attraction domain of the DFP, whose coordinates are $u^*
\approx 0.688$, $v^* \approx 0.745$, $w^* = 0$, i.e., $f_0^* \approx
1.86$, $f_2^* \approx 0.204$, $f_4^* \approx -1.43$ (here, $f_i$ are
the renormalized couplings related the Hamiltonian couplings
$f_{i,0}$). A numerical analysis indicates that the DFP can only be
reached for initial conditions that are inside the curve reported in
Fig.~\ref{flowf2f4}, i.e., for values satisfying $f_{4,0}/f_{0,0} <
g(f_{2,0}/f_{0,0}).$ Points that satisfy the opposite inequality
correspond to trajectories running away to infinity.  These results
can be rephrased in terms of $\Delta_0$. A simple calculation shows
that, if $f_{4,0}/f_{0,0} < g(f_{2,0}/f_{0,0})$, then $\Delta_0 >
0$. In other words, bicritical systems correspond to trajectories
running away to infinity: in this case the MCP corresponds to a
first-order transition. On the other hand, tetracritical systems may
show, depending on the values of the parameters, a second-order
multicritical behavior controlled by the DFP or a first-order
transition.  Note that the first-order nature of the bicritical point
should have been guessed {\em a priori}: indeed, by its very nature,
the DFP can only correspond to a tetracritical phase diagram and thus
cannot be relevant for bicritical systems.

\section{Conclusions} \label{sec4}

In conclusion, our MC simulations confirm that the O(5) FP is unstable
in the 3-$d$ O(3)$\oplus$O(2) LGW theory. The instability of the O(5)
FP is related to the presence of the spin-4 quartic perturbation in
the O(3)$\oplus$O(2) LGW Hamiltonian.  We obtain an accurate estimate
of its RG dimension, $y_{4,4}=0.23(2)$, and of the corresponding
crossover exponent, $\phi_{4,4}\equiv y_{4,4}\nu= 0.180(15)$.  
These results are slightly larger than, but in substantial agreement with, the FT
results.\cite{CPV-03} The only stable FP of the O(3)$\oplus$O(2) LGW
theory is the DFP.\cite{Aharony-02,CPV-03,PV-04} Therefore, if the
system is in its attraction domain, a decoupled critical behavior
should be observed at the MCP, with a tetracritical phase diagram.

\begin{figure}[tb]
\centerline{\psfig{width=8truecm,angle=0,file=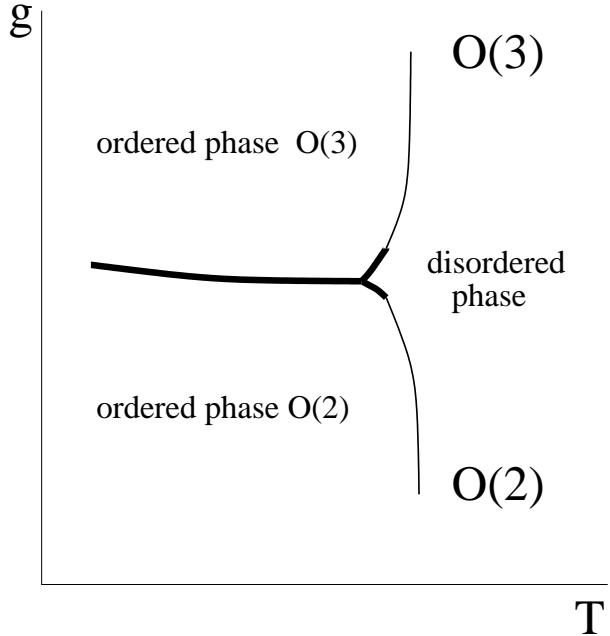}}
\vspace{2mm}
\caption{Phase diagram with a first-order bicritical point.
Here, $T$ is the temperature and $g$ 
a second relevant parameter.
The thick lines correspond to first-order transitions.
}
\label{bicr-first}
\end{figure}

As argued in Refs.~\CITE{HZ-00,DHZ-04}, sufficiently close to the MCP,
the O(3)$\oplus$O(2) LGW Hamiltonian is the effective theory of a
realistic SO(5) theory of high-$T_c$ superconductors, such as the
projected SO(5) model.\cite{ZHAHA-99,AH-00} The calculations of
Ref.~\CITE{AH-00} indicate that realistic models have a bicritical
phase diagram with $\Delta_0 < 0$.  Therefore, the analysis that we
have reported indicates that the MCP should correspond to a
first-order transition and the phase diagram should be like that
reported in Fig.~\ref{bicr-first}. Note that first-order transitions
should also be observed along the O(2) and O(3) lines close to the
MCP.  Therefore, the O(2) and O(3) lines in the temperature-doping
phase diagram are expected to present, starting from the MCP,
first-order transitions up to a tricritical point with an essentially
mean-field critical behavior (apart from logarithms), and then
second-order transitions in the O(2) and O(3) universality class,
respectively.

As put forward in Refs.~\CITE{Aharony-02,JDAHZ-03,DHZ-04}, an
effective O(5) multicritical behavior may still be observed in a
preasymptotic region when the effective breaking of the O(5) symmetry,
and in particular its spin-4 component, is small.  The observation was
essentially based on the value of the crossover exponent suggested by
an $O(\epsilon)$ calculation \cite{KNF-76},
i.e. $\phi_{4,4}=\epsilon/26$, which gives $\phi_{4,4}\simeq 0.038$
when one sets $\epsilon=1$.  Using this estimate of $\phi_{4,4}$, the
authors of Refs.~\CITE{JDAHZ-03,DHZ-04} argued that significant
deviations from the symmetric O(5) multicritical behavior cannot be
observed in experiments and realistic MC simulations, since the
crossover to the eventual asymptotic behavior is expected only when
the reduced temperature is $t_{\rm cross}\approx 10^{-10}, 10^{-11}$.
On the other hand, as shown in this paper, the actual 3-$d$ value of
$\phi_{4,4}$ is much larger. Taking for granted the arguments leading
to the estimate $t_{\rm cross} \approx 10^{-10}$, one can easily see
that, using the actual value of $\phi_{4,4}$, the crossover reduced
temperature $t_{\rm cross}$ would change from $10^{-10}$ to~\cite{FTest}
$t_{\rm cross}\approx 10^{-2}$. Taking into account that this is a
very rough estimate, which may easily miss one order of magnitude, we
conclude that the actual 3-$d$ value $\phi_{4,4}\approx 0.18$ is
sufficiently large to give rise to observable crossover effects
towards the real asymptotic behavior even in systems with a moderately
small breaking of the O(5) symmetry, such as the projected SO(5)
model.~\cite{AH-00} These crossover effects should be observable
without the need of reaching extremely small values of the reduced
temperature, as argued in Refs.~\CITE{DHZ-04,JDAHZ-03}.

We would also like to stress that the value $\phi_{4,4}\approx 0.18$
is of the same order of magnitude of the crossover exponent appearing
in other physical systems.  For instance, in randomly dilute uniaxial
magnetic materials, see, e.g., Refs.~\CITE{review,FHY-03}, the pure
Ising FP is unstable with a crossover exponent $\phi\approx 0.11$,
which is substantially smaller than $\phi_{4,4}$ at the O(5) FP.  In
these systems the asymptotic critical behavior has been precisely
observed both numerically \cite{BFMMPR-98,CMPV-03} and
experimentally,\cite{Belanger-00} and sizeable crossover effects from
the Ising to the random-exchange critical behavior have been observed
also in the case of small dilution, without the need of reaching
extremely small reduced-temperature values.

Since we have computed the initial conditions that guarantee the flow
to reach the O(5) FP, we can try to understand more quantitatively the
behavior of the model discussed in Ref.~\CITE{AH-00}. Using their
notations, at the classical level the projected SO(5) model is
equivalent to the LGW theory with
\begin{equation}
{f_{2,0}\over f_{0,0}} = {5 (\eta^2 - 1)\over 2 (2 + 3 \eta^2)}, 
\quad\qquad 
{f_{4,0}\over f_{0,0}} = 0,
\end{equation}
where $\eta$ is related to the different mobilities of the holes and
of the magnons.  The model is SO(5) invariant for $\eta = 1$ as
expected. As $\eta$ decreases (a realistic value would be
\cite{JDAHZ-03} $\eta^2 = 0.225$), $f_{2,0}/f_{0,0}$ becomes negative
and increases in absolute value. Comparing with Fig.~\ref{flowf2f4},
we see that the SO(5) breaking effects increase. An estimate of the
crossover temperature $t_{\rm cross}$ can be obtained by requiring
\cite{footnotetcross}
$\delta t^{-\phi_{4,4}} \sim 1$, where $\delta$ somehow measures the
{\em distance} between the initial condition and the line
corresponding to the O(5) invariant models. In this case, we can take
$\delta = |f_{2,0}/f_{0,0}|$ and thus estimate $t_{\rm cross}\sim
10^{-1}$ for the realistic case \cite{JDAHZ-03} $\eta^2 = 0.225$. 
Note that $t_{\rm
cross}$ decreases rapidly with increasing $\eta$ and indeed one
obtains $t_{\rm cross}\sim 10^{-3}$ for the model studied numerically
($\eta^2 = 1/2$) in Ref.~\CITE{JDAHZ-03}, thereby explaining the
apparent O(5) behavior that was observed.  Inclusion of the quantum
corrections has the effect of increasing the SO(5) breaking. For
instance, for $\eta^2 = 0.225$ we have
\begin{equation}
{f_{2,0}\over f_{0,0}} = - 0.724 - 4.36 \kappa, \qquad\qquad
{f_{4,0}\over f_{0,0}} = 1.26 \kappa,
\end{equation}
where $\kappa = {\cal V} D J I_s$ (see Ref.~\CITE{AH-00} for the definitions)
is a positive quantity. Thus, for the realistic case, the initial 
conditions are pushed even further from the line corresponding to 
the O(5) invariant models.

\appendix

\section{The five-loop series of the $\beta$ functions in the
$\overline{\rm MS}$ scheme}
\label{app1}

The expansion of the $\beta$ functions of generic
O($n_1$)$\oplus$O($n_2$) LGW theories in the modified
minimal-subtraction ($\overline{\rm MS}$) renormalization scheme is
known to five loops.\cite{CPV-03} In Ref.~\CITE{CPV-03} only the
$\epsilon$ expansions of some exponents were reported.  Here we report
the five-loop series of the $\beta$ functions of the O(3)$\oplus$O(2)
LGW theory (\ref{bicrHH}) in powers of the $\overline{\rm MS}$
renormalized quartic couplings.

In order to renormalize the O(3)$\oplus$O(2) LGW theory (\ref{bicrHH})
in the $\overline{\rm MS}$ scheme, one sets $\phi = Z_{\phi}^{1/2}
\phi_r$, $\varphi = Z_\varphi(u,v,w)^{1/2} \varphi_r$, $u_0 = A_d
\mu^\epsilon Z_{u}(u,v,w)$, $v_0 = A_d \mu^\epsilon Z_{v}(u,v,w)$,
$w_0 = A_d \mu^\epsilon Z_{w}(u,v,w)$, where $u$,$v$,$w$ are the
$\overline{\rm MS}$ renormalized quartic couplings.  The
renormalization functions $Z_{\phi,\varphi}$ and $Z_{u,v,w}$ 
are normalized so that $Z_{\phi,\varphi}\approx 1$ and $Z_{u,v,w} 
\approx u,v,w$ at tree level.  Here $A_d$ is a $d$-dependent constant 
given by $A_d\equiv 2^{d-1} \pi^{d/2} \Gamma(d/2)$.  
The $\overline{\rm MS}$ $\beta$ functions $\beta_{u,v,w}$ are 
obtained by differentiating the renormalized couplings with respect 
to the scale $\mu$, keeping the
bare couplings $u_0$, $v_0$, $w_0$ fixed.  
The O(3)$\oplus$O(2) $\beta$ functions are given by
\begin{eqnarray}
\beta_u &=& (d-4) u + \frac{11}{6} u^2 + 3 w^2 
- \frac{23}{12} u^3 - \frac{5}{2} u w^2  - 6 w^3 
+\sum_{ijk} b^{(u)}_{ijk} u^i v^j w ^k, \label{betas} \\
\beta_v &=& (d-4) v + \frac{5}{3} v^2 + \frac{9}{2} w^2 
- \frac{5}{3} v^3 - \frac{15}{4} v w^2  - 9 w^3 
+\sum_{ijk} b^{(v)}_{ijk} u^i v^j w ^k, \nonumber \\
\beta_w &=& (d-4) w + \frac{5}{6} u w + \frac{2}{3} v w + 2 w^2 
- \frac{25}{72} u^2 w - \frac{5}{2} u w^2 \nonumber\\
&&- \frac{5}{18} v^2 w  - 2 v w^2  - \frac{21}{8}w^3 
+\sum_{ijk} b^{(w)}_{ijk} u^i v^j w ^k.
\nonumber
\end{eqnarray}
The coefficients $b^{(u,v,w)}_{ijk}$ up to five loops, i.e. for $4\leq
i+j+k\leq 6$, are reported in Tables~\ref{betast}. In order to save
space, we report them numerically, although we have their exact
expressions in terms of fractions and of $\zeta$ functions with integer
argument.  
\begin{table}[tbp]
\caption{
Coefficients of the $\beta$ functions, cf. Eq.~(\ref{betas}).
}
\label{betast}
\renewcommand\arraystretch{0.4}
\begin{tabular}{cccc}
\multicolumn{1}{c}{$i,j,k$}&
\multicolumn{1}{c}{$b^{(u)}_{ijk}$}&
\multicolumn{1}{c}{$b^{(v)}_{ijk}$}&
\multicolumn{1}{c}{$b^{(w)}_{ijk}$}\\
\tableline \hline
4,0,0 & 5.95643    & 0          & 0         \\
3,1,0 & 0          & 0          & 0         \\
3,0,1 & 0          & 0          & 0.902778  \\
2,2,0 & 0          & 0          & 0          \\
2,1,1 & 0          & 0          & 0          \\
2,0,2 & 2.21875    & 0.234375   & 4.44611    \\
1,3,0 & 0          & 0          & 0          \\
1,2,1 & 0          & 0          & 0           \\
1,1,2 & $-$1.08333 & $-$2.03125 & 1.38889   \\
1,0,3 &  38.6747   & 15         & 14.0051   \\
0,4,0 & 0          & 4.99347    & 0         \\
0,3,1 & 0          & 0          & 0.652778  \\
0,2,2 & 0.125      & 3.46875    & 3.50134   \\
0,1,3 & 8          & 54.262     & 11.6624   \\
0,0,4 & 13.9123    & 20.4465    & 12.8874   \\
\hline
5,0,0 & $-$27.3529 &  0         & 0         \\
4,1,0 & 0          &  0         & 0         \\
4,0,1 & 0          &  0         & $-$2.91484 \\
3,2,0 & 0          &  0         & 0 \\
3,1,1 & 0          &  0         & 0\\
3,0,2 & $-$5.56225 &$-$1.0098   &$-$17.0442 \\
2,3,0 & 0          &  0         & 0 \\
2,2,1 & 0          &  0         & 0 \\
2,1,2 & 1.18623    &$-$0.0839266&$-$2.56238 \\ 
2,0,3 & $-$188.898 &$-$21.3002  &$-$51.1747 \\
1,4,0 & 0          &  0         & 0 \\
1,3,1 & 0          &  0         & 0 \\
1,2,2 & $-$0.144409& 1.81816    &$-$2.57945 \\
1,1,3 & $-$19.1406 &$-$34.1777  &$-$41.1293 \\
1,0,4 & $-$173.188 &$-$169.835  &$-$108.727 \\
0,5,0 & 0          &$-$21.9072  & 0 \\
0,4,1 & 0          &  0         &$-$1.99192 \\ 
0,3,2 & $-$0.489598&$-$7.99393  &$-$12.3167 \\ 
0,2,3 & $-$10.9914 &$-$247.731  &$-$39.1403 \\
0,1,4 & $-$91.5308 &$-$ 239.138 &$-$94.7246 \\
0,0,5 & $-$120.043 &$-$191.964  &$-$100.117 \\
\hline
6,0,0 & 156.207   &  0          & 0  \\
5,1,0 & 0         &  0          & 0 \\
5,0,1 & 0         &  0          & 12.3408 \\
4,2,0 & 0         &  0          & 0\\
4,1,1 & 0         &  0          & 0 \\
4,0,2 & 6.47354   & 2.81164     & 80.795 \\
3,3,0 & 0         & 0           & 0 \\
3,2,1 & 0         & 0           & 0 \\
3,1,2 & 2.43996   &$-$0.068611 & 11.0339 \\
3,0,3 & 1146.82   & 77.9976     & 249.36 \\
2,4,0 & 0         & 0           & 0 \\
2,3,1 & 0         & 0           &  0 \\
2,2,2 & 1.98376   & 3.09444     & 5.96446 \\ 
2,1,3 & 35.922    & 57.0315     & 124.802 \\
2,0,4 & 1294.73   & 518.911     & 617.85 \\
1,5,0 & 0         & 0           & 0 \\
1,4,1 & 0         & 0           & 0 \\
1,3,2 &$-$0.142137& 4.13976     & 9.84045 \\
1,2,3 & 34.0863   &  66.1707    & 119.023 \\
1,1,4 & 659.256   & 1136.61     & 636.697 \\
1,0,5 & 1776.58   & 1751.3      & 1155.64 \\
0,6,0 & 0         & 120.141     & 0 \\
0,5,1 & 0         & 0           & 7.99517 \\
0,4,2 & 1.27762   & 10.0207     &  54.8793 \\
0,3,3 & 37.2454   & 1426.49     & 178.161 \\
0,2,4 &  257.6    & 1657.51     & 484.147 \\
0,1,5 & 943.812   & 2594.22     & 1031.67 \\
0,0,6 & 1108.93   & 1879.79     &852.872 \\
\end{tabular}
\end{table}


\begin{references}
\bibitem{FTest}
Using the slightly smaller FT estimate $\phi_{4,4}\approx 0.15$,
the change would be from $t_{\rm cross}\approx 10^{-10}$ to
$t_{\rm cross}\approx 2 \times 10^{-3}$ (Refs.~\CITE{JDAHZ-03,CPV-03}).
\bibitem{footnotetcross}
We should warn that this is just a rough estimate, which may easily
miss one order of magnitude. A more reliable estimate of $t_{\rm
cross}$ would require the determination of the scaling free energy
(\ref{freen}) for $g_2=0$, $\bar{f}(y)\equiv f(0,y)$.  Around $y=0$,
we may expand it as $\bar{f}(y) = \bar{f}_0 + \bar{f}_1 y + O(y^2)$.
Once fixed the normalizations, one may estimate $t_{\rm cross}$ from
$g_4 t_{\rm cross}^{-\phi_{4,4}}\approx \bar{f}_0/\bar{f}_1$.
\end{references}
\end{document}